# A SECURE COLOR IMAGE STEGANOGRAPHY IN TRANSFORM DOMAIN

Hemalatha S[1], U Dinesh Acharya[2], Renuka A[3], Priya R. Kamath[4]

[1,2,3,4]Department of Computer Science and Engineering, Manipal Institute of Technology, Manipal University, Manipal, Karnataka, India
[1]hema.shama@manipal.edu;[2]dinesh.acharya@manipal.edu;[3]renuka.prabhu@manipal.edu;[4]priyarkamath@gmail.com

*ABSTRACT*

*Steganography is the art and science of covert communication. The secret information can be concealed in content such as image, audio, or video. This paper provides a novel image steganography technique to hide both image and key in color cover image using Discrete Wavelet Transform (DWT) and Integer Wavelet Transform (IWT). There is no visual difference between the stego image and the cover image. The extracted image is also similar to the secret image. This is proved by the high PSNR (Peak Signal to Noise Ratio), value for both stego and extracted secret image. The results are compared with the results of similar techniques and it is found that the proposed technique is simple and gives better PSNR values than others.*

## 1. INTRODUCTION

Recently, the information hiding technique has developed rapidly in the field of information security and has received significant attention from both industry and academia. It contains two main branches: digital watermarking and steganography. The former is mainly used for copyright protection of electronic products while, the latter is a way of covert communication. The main purpose of steganography is to convey the information secretly by concealing the very existence of information in some other medium such as image, audio or video. The content used to embed information is called as cover object. The cover along with the hidden information is called as stego-object [1]. In this paper image is the cover and secret information is also an image. Both secret image and stego key are embedded in the cover image to get stego image.

The major objective of steganography is to prevent some unintended observer from stealing or destroying the confidential information. There are some factors to be considered when designing a steganography system: [1]

- Invisibility: Invisibility is the ability to be unnoticed by the human.
- Security: Even if an attacker realizes the existence of the information in the stego object it should be impossible for the attacker to detect the information. The closer the stego image to the cover image, the higher the security. It is measured in terms of PSNR. High PSNR value indicates high security.

PSNR = $10 \log \frac{L^2}{\sqrt{MSE}}$ dB, where L = maximum value, MSE = Mean Square Error.

MSE = $\frac{1}{N}\sum_{i=1}^{N}|Xi - Xi'|$, where X = original value, X' = stego value and N = number of samples.

- Capacity: The amount of information that can be hidden relative to the size of the cover object without deteriorating the quality of the cover object
- Robustness: It is the ability of the stego to withstand manipulations such as filtering, cropping, rotation, compression etc.

The design of a steganographic system can be categorized into spatial domain methods and transform domain methods [1].

In spatial domain methods, the processing is applied on the image pixel values directly. The advantage of these methods is simplicity. The disadvantage is low ability to bear signal processing operations. Least Significant Bit Insertion methods, Pallete based methods come under this category.

In transform domain methods, the first step is to transform the cover image into different domain. Then the transformed coefficients are processed to hide the secret information. These changed coefficients are transformed back into spatial domain to get stego image. The advantage of transform domain methods is the high ability to face signal processing operations. However, methods of this type are computationally complex. Steganography methods using DCT (Discrete Cosine Transforms), DWT, DFT (Discrete Fourier Transforms) come under this category.

## 1.1. Discrete Wavelet Transform

DWT is used for digital images. Many DWTs are available. Depending on the application appropriate one should be used. The simplest one is haar transform. To hide text message integer wavelet transform can be used. When DWT is applied to an image it is decomposed into 4 sub bands: LL, HL, LH and HH. LL part contains the most significant features. So if the information is hidden in LL part the stego image can withstand compression or other manipulations. But sometimes distortion may be produced in the stego image and then other sub bands can be used [1]. The decomposition of Lena image by 2 levels of 2D - DWT is shown in Figure 1.

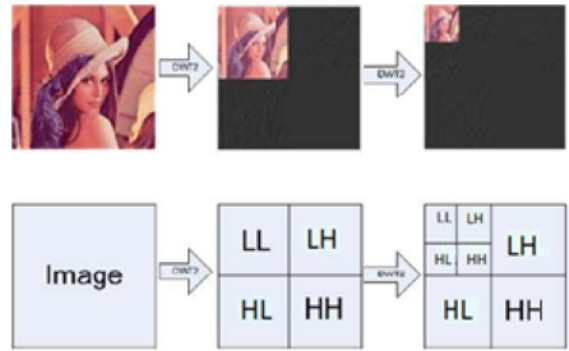

Figure 1. 2 Level 2D – DWT

## 1.2. Integer Wavelet Transform

IWT is a more efficient approach to lossless compression. The coefficients in this transform are represented by finite precision numbers which allows for lossless encoding. This wavelet transform maps integers to integers. In case of DWT, if the input consists of integers (as in the case of images), the resulting output no longer consists of integers. Thus the perfect reconstruction of the original image becomes difficult. However, with the introduction of Wavelet transforms that map integers to integers the output can be completely characterized with integers. The LL sub-band in the case of IWT appears to be a close copy with smaller scale of the original image while in the case of DWT the resulting LL sub-band is distorted slightly, as shown in Figure 2.[2].

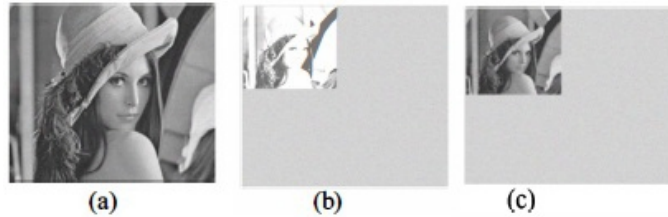

Figure 2 (a) Original image Lena. (b) One level DWT in sub band LL (c) One level IWT in sub-band LL.

If the original image (I) is X pixels high and Y pixels wide, the level of each of the pixel at (i,j) is denoted by $I_{i,j}$.[3]

The IWT coefficients are given by

$$LL_{i,j} = \lfloor ( I_{2i, 2j} + I_{2i+1, 2j}) /2 \rfloor \quad (1)$$
$$HL_{i,j} = I_{2i+1, 2j} - I_{2i, 2j} \quad (2)$$
$$LH_{i,j} = I_{2i, 2j+1} - I_{2i, 2j} \quad (3)$$
$$HH_{i,j} = I_{2i+1, 2j +1} - I_{2i, 2j} \quad (4)$$

The inverse transform is given by

$$I_{2i, 2j} = LL_{i,j} - \lfloor HL_{i,j}/2 \rfloor \quad (5)$$
$$I_{2i, 2j +1} = LL_{i,j} + \lfloor (HL_{i,j+1})/2 \rfloor \quad (6)$$
$$I_{2i+1, 2j} = I_{2i, 2j +1} + LH_{i,j} - HL_{i,j} \quad (7)$$
$$I_{2i+1, 2j+1} = I_{2i+1, 2j} + HH_{i,j} - LH_{i,j} \quad (8)$$

where, $1 \leq i \leq X/2$, $1 \leq j \leq Y/2$ and $\lfloor \rfloor$ denotes floor value.

## 2. RELATED WORK

Color images are represented in different color spaces such as RGB (Red Green Blue), HSV (Hue, Saturation, Value), YUV, YIQ, YCbCr (Luminance/Chrominance) etc. YCbCr is one of the best representations for steganography because the eye is sensitive to small changes in luminance but not in chrominance, so the chrominance part can be altered, without visually impairing the overall image quality much. Y is luminance component and CbCr are the blue and red chrominance components respectively. The values in one color space can be easily converted into another color space using conversion formula [4].

S. M. Masud Karim, et al., [5] proposed a new approach based on LSB using secret key. The secret key encrypts the hidden information and then it is stored into different position of LSB of image. This provides very good security. XIE Qing et al.,[6] proposed a method in which the information is hidden in all RGB planes based on HVS (Human Visual System). This degrades the quality of the stego image. In the method proposed by Sunny Sachdeva et al., [7] the Vector Quantization (VQ) table is used to hide the secret message which increases the capacity and also stego size. Sankar Roy et al., [8] proposed an improved steganography approach for hiding text

19

International Journal on Cryptography and Information Security (IJCIS), Vol.3, No.1, March 2013

messages within lossless RGB images which will suffer from withstanding the signal processing operations. Minimum deviation of fidelity based data embedding technique has been proposed by J. K. Mandal et al, [9] where two bits per byte have been replaced by choosing the position randomly between LSB and up to fourth bit towards MSB. A DWT based frequency domain steganographic technique, termed as WTSIC is also proposed by the same authors, [10] where secret message/image bits stream are embedded in horizontal, vertical and diagonal components. Anjali Sejul, et al, [4] proposed an algorithm in which binary images are considered to be secret images which are embedded inside the cover image by taking the HSV (Hue, Saturation, Value) values of the cover image into consideration. The secret image is inserted into the cover image by cropping the cover image according to the skin tone detection and then applying the DWT. In this method the capacity is too low.

El Safy et.al, [11], used an adaptive steganographic technique based on IWT, which improves the hiding capacity and PSNR. Neda Raftari and Amir Masoud E. M. [12] used IWT and Munkres' assignment algorithm which embeds secret image in frequency domain of cover image with high matching quality. The improvement is obtained with higher computation. Saddaf Rubab et al.,[13] proposed a complex method using DWT and Blowfish encryption technique to hide text message in color image. In the paper by Kapre Bhagyashri et al, [14] a new singular value decomposition (SVD) and DWT based water mark technique is proposed in full frequency band in YUV color space. Nabin Ghoshal et al., uses a steganographic scheme for colour image authentication (SSCIA) [15] where the watermark image is embedded using DFT.

The proposed work is the extension of our previous work [16] to color images in which the secret image is transmitted without actually embedding in the cover image. Only the key is hidden in the cover image. The steps for embedding are as follows:

- Obtain single level 2D DWT of the cover-image C and secret-image S.
- The resulting transformed matrix consists of four sub-bands CLL, CHL, CLH and CHH and SLL, SHL, SLH and SHH obtained by transforming images C and S respectively.
- The sub-images CLL and SLL are subdivided into non-overlapping blocks $BC_{k1}$ ($1 \leq k1 < nc$) and $BS_i$ ($1 \leq i < ns$) of size 2x2 where nc, ns are the total number of non-overlapping blocks obtained from sub-images CLL and SLL respectively.
- Every block $BS_i$, is compared with block $BC_{k1}$. The pair of blocks which have the least Root Mean Square Error is determined. A key is used to determine the address of the best matched block $BC_{k1}$ for the block $BS_i$. Then inverse 2D DWT is applied to obtain C.
- The Key is then stored using one of the spatial domain techniques in the cover image C. The simplest of the spatial domain techniques is LSB insertion algorithm.
- The resultant image is a stego-image.
- 

The secret image can now be extracted from this image by following the steps mentioned below:

- From the stego-image G, obtain the secret key K1
- Transform the stego-image into single level 2D DWT.
- This transformation results in four sub-bands GLL, GHL, GLH and GHH.
- Divide the sub-band image GLL into 2x2 non-overlapping blocks. The secret key K1 is used to obtain the blocks that have the nearest approximation to the original blocks in secret image.
- The obtained blocks are then rearranged to obtain the sub-band image SLLnew. Assuming SHLnew, SLHnew, SHHnew are zero matrices of dimension similar to SLLnew , 2D IDWT (Inverse DWT) is obtained.
- The resultant image is the secret image that was originally intended to be embedded within the cover-image.

## 3. PROPOSED METHOD

In the proposed method, the cover is 256x256 lena color image. The secret information is grey scale image of size 128 x128. To transfer the secret image confidentially, the secret image itself is not hidden, instead a key is generated and then the key is encrypted and Run Length Encoded. The resultant key is hidden in the cover image using Integer Wavelet Transform (IWT). This improves the security and also the capacity can be improved to some extent since the key is compressed.

### 3.1. Key Generation

To generate the key following steps are performed.

- Represent the cover image C in YCbCr color space
- Obtain single level 2D DWT of secret-image S and Cr component of C.
- The resulting transformed matrix consists of four sub-bands SLL, SHL, SLH and SHH and CLL, CHL, CLH and CHH obtained by transforming S and Cr component of C respectively. After this the steps are same as our previous work, which are repeated again for better understanding.
- The sub-images CLL and SLL are subdivided into non-overlapping blocks $BC_{k1}$ (1 $\leq$ k1 < nc) and $BS_i$ (1 $\leq$ i < ns) of size 2x2 where nc, ns are the total number of non-overlapping blocks obtained from sub-images CLL and SLL respectively.
- Every block $BS_i$, is compared with block $BC_{k1}$. The pair of blocks which have the least Root Mean Square Error is determined. A key is used to determine the address of the best matched block $BC_{k1}$ for the block $BS_i$. Then inverse 2D DWT is applied to obtain Cr component.
- The key is then encrypted using simple exclusive or operation with a key and run length encoded.

### 3.2. Key Embedding

The key obtained in the previous subsection is hidden in the cover image using IWT. The steps are as follows:

- Find the integer wavelet transform of Cr component of the cover image.
- Replace the least significant bit planes of the higher frequency components of the transformed image by the bits of the key.
- Obtain the inverse IWT of the resulting image to get the stego Cr component.
- Represent the resultant image in RGB color space to obtain stego image G.

The secret image can now be extracted from this image using the following steps:

### 3.3. Key Extraction

The steps are as follows:

- Represent the stego image G in YCbCr color space.
- Find the integer wavelet transform of Cr component of the stego image G
- Obtain the key from the least significant bit planes of the higher frequency components of the transformed image. Convert back to RGB representation.
- Decompress the key and then decrypt it to get original key.

### 3.4. Secret Image Generation

To generate the secret image following steps are performed

- Transform the stego-image G into single level 2D DWT.
- This transformation results in four sub-bands GLL, GHL, GLH and GHH.
- Divide the sub-band image GLL into 2x2 non-overlapping blocks. The key is used to obtain the blocks that have the nearest approximation to the original blocks in secret image.
- The obtained blocks are then rearranged to obtain the sub-band image SLLnew. Assuming SHLnew, SLHnew, SHHnew are zero matrices of dimension similar to SLLnew, 2D IDWT is obtained.
- The resultant image is the secret image S.

## 4. EXPERIMENTAL RESULTS

The algorithm is tested in MATLAB with Cr component of the cover image. The results with different cover images and secret images are shown. Original cover and secret images are shown in Figure 3 and Figure 4 respectively. Football is hidden in lena, earth is hidden in peppers and moon is hidden in baboon. The cover image size is 256x256 and secret image size is 128x128.

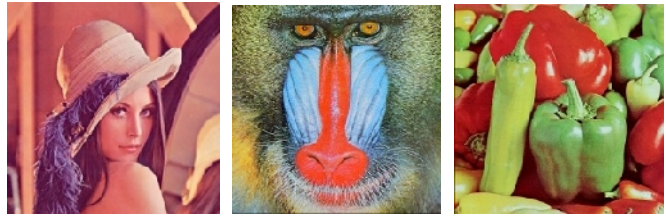

Figure 3. Color images that are used as cover images: (a) lena   (b) baboon   (c) peppers

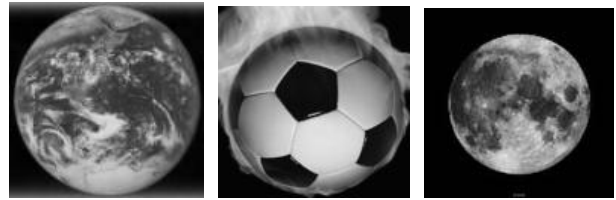

Figure 4. Images which are used as secret images: (a) earth   (b) football   (c) moon

The stego and extracted secret images are shown in Figure 5 and Figure 6 respectively.

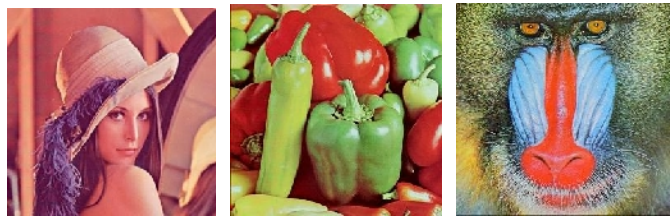

Figure 5. Stego images: (a) football, (b) earth, (c) moon as secret images

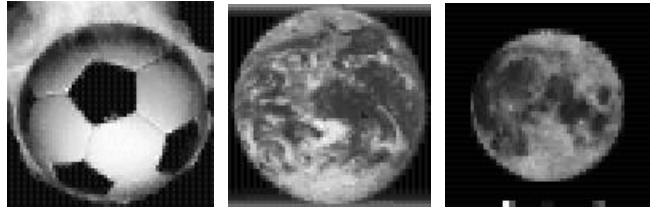

Figure 6. Extracted Secret images: (a) football   (b) earth   (c) moon

The PSNR in dB in all cases for stego and extracted secret images are tabulated in Tables 1 and 2 respectively.

Table 1.   PSNR (in dB) of the stego image

| COVER IMAGE (256x256) | SECRET IMAGE(128x128) | | |
|---|---|---|---|
| | football | earth | moon |
| lena | 44.3 | 44.4 | 44.2 |
| peppers | 44.7 | 44.7 | 45.0 |
| baboon | 44.8 | 44.8 | 45.0 |

Table 2.   PSNR (in dB) of the  extracted secret image

| COVER IMAGE (256x256) | SECRET IMAGE(128x128) | | |
|---|---|---|---|
| | football | earth | moon |
| lena | 37.5 | 34.1 | 30.7 |
| peppers | 30.4 | 28.6 | 26.7 |
| baboon | 37.2 | 27.6 | 36.5 |

Table 3 compares the PSNR values in the proposed method and that in the other four methods. In all these the cover image considered is lena image and the secret images used are of comparable sizes.

Table 3.  Comparison of  PSNR (in dB) of the stego image in different methods

| TECHNIQUE | PSNR |
|---|---|
| Mandal, J.K. et al. [9] | 39.6 |
| Mandal, J.K. et al. [10] | 42.4 |
| Kapre Bhagyashri, S. et al. [14] | 36.6 |
| Ghoshal, N. et al. [15] | 33.2 |
| PROPOSED | 44.3 |

## 5. CONCLUSIONS

In this paper, we observe that the secret image can be regenerated without actually storing the image itself. This approach results in high quality of the stego-image having high PSNR values compared to other methods. Instead of taking the least significant bit plane to hide the key middle bit planes can be considered to improve the security. One assumption made in this paper is that the encryption key is sent to the recepient by some means. Standard encryption techniques like Blowfish or RC6 can be used to encrypt the key so that the security can be further improved. The encryption key also has to be hidden in the cover image.